\pdfoutput=1

\documentclass[11pt,twoside,a4paper,cmspaper,final,collab]{cms-tdr}
\def\svnVersion{167115}\def\cmsCernNo{2012-150}\def\cmsCernDate{\today}\def\cmsMessage{Submitted to Physical Review Letters}

\begin{document}\cmsNoteHeader{EXO-11-041}

\hyphenation{had-ron-i-za-tion}
\hyphenation{cal-or-i-me-ter}
\hyphenation{de-vices}

\RCS$Revision: 144562 $
\RCS$HeadURL: svn+ssh://svn.cern.ch/reps/tdr2/papers/EXO-11-041/trunk/EXO-11-041.tex $
\RCS$Id: EXO-11-041.tex 144562 2012-08-22 16:59:06Z alverson $
\newlength\cmsFigWidth
\ifthenelse{\boolean{cms@external}}{\setlength\cmsFigWidth{0.85\columnwidth}}{\setlength\cmsFigWidth{0.4\textwidth}}
\ifthenelse{\boolean{cms@external}}{\providecommand{\cmsLeft}{top}}{\providecommand{\cmsLeft}{left}}
\ifthenelse{\boolean{cms@external}}{\providecommand{\cmsRight}{bottom}}{\providecommand{\cmsRight}{right}}

\newcommand{\rhoT}{\ensuremath{\rho_\mathrm{TC}}\xspace}
\newcommand{\aT}{\ensuremath{\cmsSymbolFace{a}_\mathrm{TC}}\xspace}
\newcommand{\omegaT}{\ensuremath{\omega_\mathrm{TC}}\xspace}
\newcommand{\piT}{\ensuremath{\Pgp_\mathrm{TC}}\xspace}

\newcommand{\usedLumiWZ}{\ensuremath{4.98 \pm 0.11\fbinv}\xspace}
\newcommand{\wprimeLimit}{1143}
\newcommand{\tcLesHouchesLimit}{687}
\newcommand{\tcLesHouchesLimitLower}{167}
\newcommand{\tcOptimisticLimitObs}{938}
\newcommand{\tcOptimisticLimitObsLower}{180}

\cmsNoteHeader{EXO-11-041} 

\title{Search for a \texorpdfstring{\PWpr\ or $\rho_\mathrm{TC}$ decaying to $\PW\cPZ$ in $\Pp\Pp$ collisions at $\sqrt{s}=7$\TeV}{W' or technirho decaying to WZ in pp collisions at sqrt(s) = 7 TeV}}

\date{\today}

\abstract{
	A search is performed in $\Pp\Pp$ collisions at $\sqrt{s}=7$\TeV for exotic particles decaying via $\PW\cPZ$ to final states with electrons and muons. The data sample corresponds to an integrated luminosity of approximately 5\fbinv. No significant excess is observed in the data above the
	expected standard model background. Upper bounds at 95\% confidence level are set on the production cross section
	of the \PWpr\ boson described by the sequential standard model, and on the $\PWpr\PW\cPZ$ coupling.
	$\PWpr$ bosons with  masses below 1143\GeV are excluded. Limits are also set in the context of low-scale technicolor models, under a range of assumptions concerning the model parameters.
}

\hypersetup{%
pdfauthor={CMS Collaboration},%
pdftitle={Search for a Wprime or techni-rho decaying to WZ in pp collisions at sqrt(s)=7 TeV},%
pdfsubject={CMS},%
pdfkeywords={CMS, physics}}

\maketitle 

The standard model (SM) of particle physics has passed many rigorous tests and its predictions have often been matched by experimental data with amazing precision. However, it is widely accepted that the SM cannot be the ultimate theory of fundamental particles and their interactions since it has a number of shortcomings, e.g., it fails to incorporate gravity and has no explanation for the dominance of matter over anti-matter in the universe. Various extensions of the SM have been proposed to address these problems and to explain the mechanism of electroweak symmetry breaking. Many of these models predict the existence of a new heavy charged gauge boson, generically known as $\PWpr$, that decays into a pair of $\PW$ and \cPZ\ bosons~\cite{ssm1,ssm3,Birkedal:2004au,littleHiggs,Agashe:2008jb,Grojean:2011vu}. Previous $\PWpr$  searches have typically interpreted
their results in terms of the sequential standard model
(SSM)~\cite{AtlasLNu2011,CMSLNu2012,Abazov:2009eu,PhysRevLett.104.241801,PhysRevLett.107.011801,AtlasWZ2012}, a simple extension of the SM in
which the couplings of the $\PWpr$  to fermions are identical to those of the \PW. Many of these searches have been conducted in leptonic final states and assume that the $\PWpr \rightarrow \PW\cPZ$ decay mode is suppressed. Searches for exotic particles that decay into  $\PW\cPZ$  pairs are thus complementary to searches in the leptonic channels. Moreover, there are other models in which the $\PWpr$  couplings to SM fermions are suppressed, giving rise to a fermiophobic $\PWpr$  with an enhanced coupling to $\PW$ and $\cPZ$ bosons~\cite{Wprimereview, PhysRevD.78.031701}. It is therefore important to search for $\PWpr$  bosons also in the $\PW\cPZ$ final state.

Another model predicting a new heavy boson decaying into $\PW\cPZ$ is technicolor (TC):  a gauge theory modeled on QCD with no elementary scalar particles~\cite{Weinberg:1979bn,Susskind:1978ms}. TC provides a dynamical explanation of electroweak symmetry breaking by generating masses of the $\PW$ and $\cPZ$ bosons through the binding energy of techni-fermions. Furthermore, it predicts a series of techni-hadrons that are bound states of the new strong interaction. By analogy with QCD, the techni-hadrons with
$I^G(J^{PC}) = 1^-(0^{-+})$, $1^+(1^{--})$, and $1^-(1^{++})$ are called
$\piT$, $\rhoT$, and $\aT$ respectively. In low-scale technicolor
(LSTC)~\cite{tc2,tc1} the lightest techni-hadrons are expected to have
masses below 700\GeV  with the charged $\rhoT$ and $\aT$ able to decay to
$\PW\cPZ$ boson pairs. Since these two states are expected to be nearly mass-degenerate~\cite{tc1}, they would appear as a single feature in the $\PW\cPZ$ invariant mass
spectrum, and we hereafter refer to them collectively as $\rhoT$. The
relationship between the masses of \rhoT and \piT, $M(\rhoT)$ and $M(\piT)$, significantly affects the
$\rhoT$ branching fractions~\cite{LesHouches09}. If $M(\rhoT)<2M(\piT)$, the decay
$\rhoT \to \piT + \PW$ dominates, such that the branching fraction $\mathcal{B}(\rhoT \to \PW\cPZ)<10\%$. However, if this decay is kinematically inaccessible,
$\mathcal{B}(\rhoT \to \PW\cPZ)$ approaches 100\%.

This Letter presents a search for new particles decaying via a $\PW\cPZ$ pair with $\PW\to\ell\Pgn$ and $\cPZ\to\ell\ell$ in the final state, where $\ell= \Pe, \Pgm$. The results are interpreted in the context of a SSM $\PWpr$ boson and a LSTC $\rhoT$ particle. A previous search in this channel performed by the \DZERO experiment excludes $\PWpr$ bosons with masses between 188 and 520\GeV at 95\% confidence level (CL)~\cite{Abazov:2009eu}. Their result also excludes $\rhoT$ between 208 and 409\GeV at 95\% CL, under the assumption that $M(\rhoT) < M(\piT) + M(\PW)$. The analysis presented here considers the case where the relations between parameters are those of Ref.~\cite{LesHouches09}, $M(\piT)=\frac{3}{4}M(\rhoT) - 25\GeV$, and also investigates the results of varying  the $\rhoT$ and $\piT$ masses.

\label{sec:CMS}
This study uses data corresponding to an integrated luminosity of \usedLumiWZ of proton-proton collisions at $\sqrt{s}=7\TeV$,  recorded by the Compact Muon Solenoid (CMS) experiment at the Large Hadron Collider (LHC) in 2011.
The central feature of the apparatus is a superconducting solenoid, of 6\unit{m} internal diameter, providing a field of 3.8\unit{T}. Inside the magnet coil are the silicon pixel and strip tracker, the lead tungstate crystal electromagnetic calorimeter, and the brass-scintillator hadron calorimeter. Muons are measured in gas-ionization detectors embedded in the steel return yoke. In addition to the barrel and endcap detectors, CMS has extensive forward calorimetry.
The trigger system, composed of a custom hardware layer feeding into a commercial processor farm, reduces the event rate to approximately 300\unit{Hz} for storage and further analysis. A detailed description of the CMS apparatus may be found elsewhere~\cite{:2008zzk}.

The $\PW\cPZ \to 3\ell\Pgn$ decay under study is characterized by a pair of same-flavor, opposite-charge,
isolated leptons with high transverse momentum (\pt), having an invariant mass consistent with that of the $\cPZ$ boson, along with a third, high-\pt, isolated lepton, and missing transverse energy (\MET) associated with the escaping neutrino.
Other sources of events with three leptons, genuine or misidentified, constitute the background, and can be grouped into the following classes:
\begin{itemize}
\item The irreducible SM $\PW\cPZ$ background.
\item Non-resonant events with no genuine $\cPZ$ boson in the final state, including top pair ($\ttbar$), multi-jet, \PW+jet, $\PW\gamma$ and \PW\PW+jet production. Only the first
	of these makes a significant background contribution, and the others are therefore not considered in this analysis.
\item Events with a genuine $\cPZ$ boson decaying to leptons and a third misidentified or non-isolated lepton that is reconstructed as isolated. These events include  \cPZ+jets (both light and heavy flavor) and $\cPZ\gamma$ processes.

\item Events with a genuine $\cPZ$ boson decaying to leptons and a third genuine isolated lepton, dominated by $\cPZ\cPZ \to 4\ell$ decays in which one of the four
	leptons is undetected. Although irreducible, this contribution is small because of the low $\cPZ\cPZ$ production cross section and dilepton decay branching fraction.
\end{itemize}

The background was modeled using samples produced with a full \GEANTfour~\cite{GEANT} simulation of the CMS detector. The primary SM background arises from the
$\PW\cPZ \to 3\ell\Pgn$ process, which was generated using the \MADGRAPH 5.1~\cite{Alwall:2007st} generator, interfaced to \PYTHIA~6.4.22~\cite{Sjostrand:2006za} for parton showering, hadronization, and simulation of the underlying event. The CTEQ6L1~\cite{Pumplin:2002vw}
parton distribution functions (PDF) were used with \PYTHIA tune Z2~\cite{UEPAS}.
Higher-order effects were estimated using next-to-leading-order (NLO) K-factor corrections, obtained using \MCFM~6.1~\cite{MCFM}. The other background processes were also generated with \MADGRAPH in combination with \PYTHIA, with the exception of the $\cPZ\cPZ$ process, which was generated using \POWHEG~1.1~\cite{Frixione:2007vw}. The signal was simulated using \PYTHIA~6.4.22 with mass-dependent next-to-next-to-leading-order (NNLO) cross sections obtained using the simulation code \textsc{fewz}~2.0~\cite{Gavin:2010az}. Characteristic signal widths are in the range 50-150\GeV for the $\PWpr$ and 50-70\GeV for the $\rhoT$ masses examined. These are dominated by the detector resolution as the natural widths are approximately 10 and 2 \GeV, respectively.

Candidate events were triggered using a double-electron or double-muon requirement, with \pt thresholds of 17\GeV and 8\GeV respectively for the highest-\pt and second-highest-\pt leptons.
In the offline selection, events were required to have at least three reconstructed leptons within the tracking acceptance of $|\eta| < 2.5$ (2.4) for electrons (muons),
where $\eta \equiv -\ln(\tan(\theta/2))$, and $\theta$ is the polar angle with respect to  the counterclockwise proton beam. To reduce background from jets
misidentified as leptons, all lepton candidates were required to satisfy a series of identification and isolation criteria. In calculating isolation variables,
the track momenta and energy deposits, excluding those associated with the lepton itself, were summed in a cone of $\Delta R < 0.3$ around the lepton direction, where $\Delta R = \sqrt{(\Delta \phi)^2 + (\Delta \eta)^2}$, and divided by the lepton transverse momentum. These sums were corrected for additional proton-proton interactions in each bunch crossing (pile-up) using
the \textsc{fastjet} energy density technique~\cite{Fastjet1, Fastjet2}. For simulated samples, pile-up was modeled by superimposing generated minimum-bias interactions
onto simulated events, weighted such that the interaction multiplicity agreed with the luminosity profile of the data set used. An additional scale factor (equal to one within 5\%) derived
from ``tag-and-probe''~\cite{CMS:2011aa} studies was applied to simulated events to correct for differences in lepton selection efficiency measured in recorded and simulated event samples.

\cPZ\ boson candidates were reconstructed from pairs of opposite-sign, same-flavor leptons with the highest and second-highest lepton \pt greater than 20 and 10\GeV, respectively,
and with an invariant mass between 60 and 120\GeV. In events where more than one such pair was found, the one with invariant mass closest to the nominal $\cPZ$ mass was selected.
If four leptons compatible with two distinct $\cPZ$ candidates were present, the event was rejected in order to suppress $\cPZ\cPZ$ background. The candidate for the \PW-boson decay
product was required to pass tighter isolation and identification requirements. If multiple lepton candidates existed, the highest-\pt remaining lepton, with \pt greater
than 20\GeV, was chosen. Finally, candidate events were required to have $\MET > 30\GeV$, as measured with a particle-flow algorithm~\cite{pf}, in order to discriminate
against Z+jets events with high-\pt jets misidentified as leptons, and against $\cPZ\gamma$ events with converted photons.

\begin{figure}[htb]
 \begin{center}
 \includegraphics[width=0.45\textwidth]{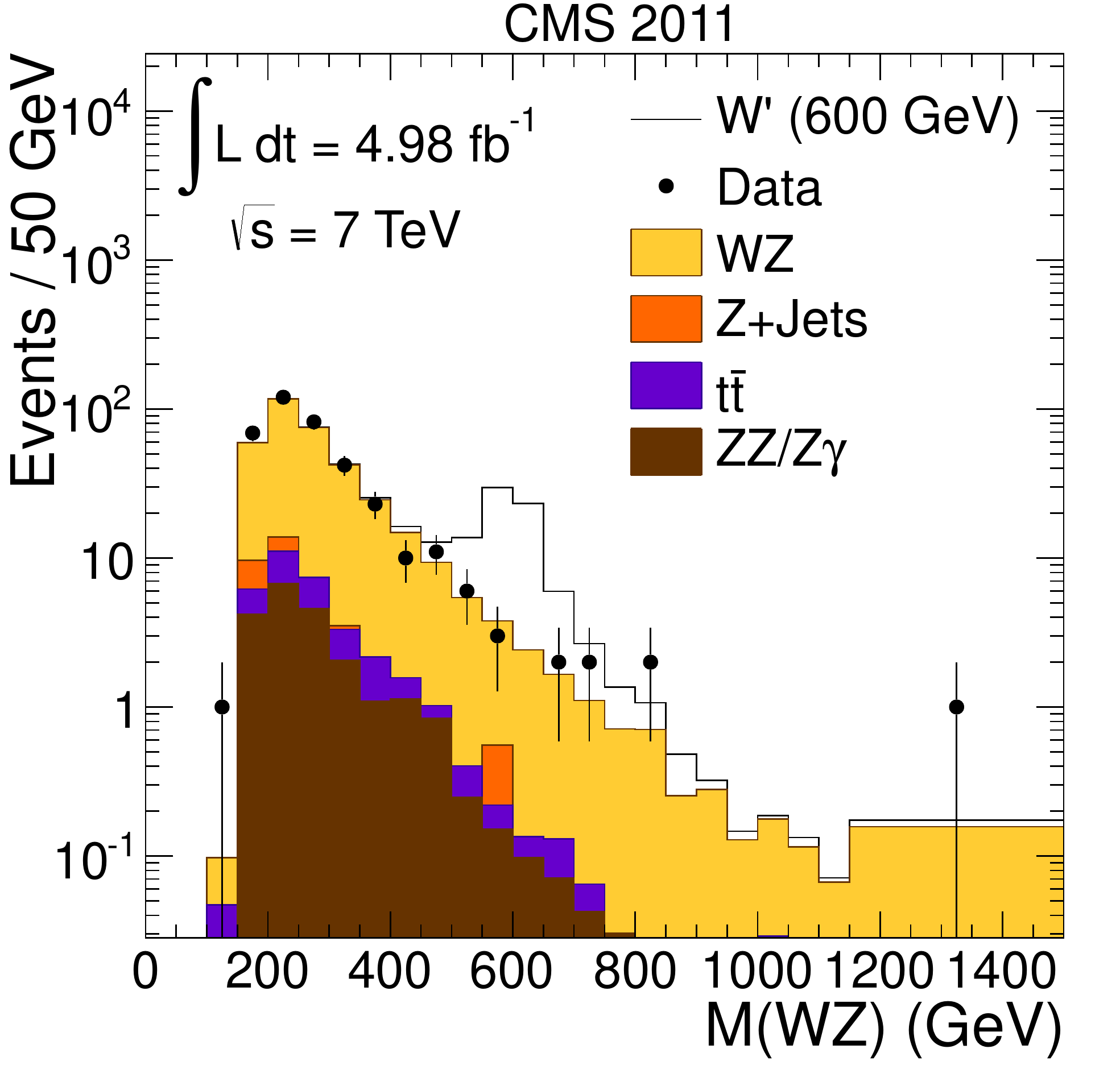}
 \includegraphics[width=0.45\textwidth]{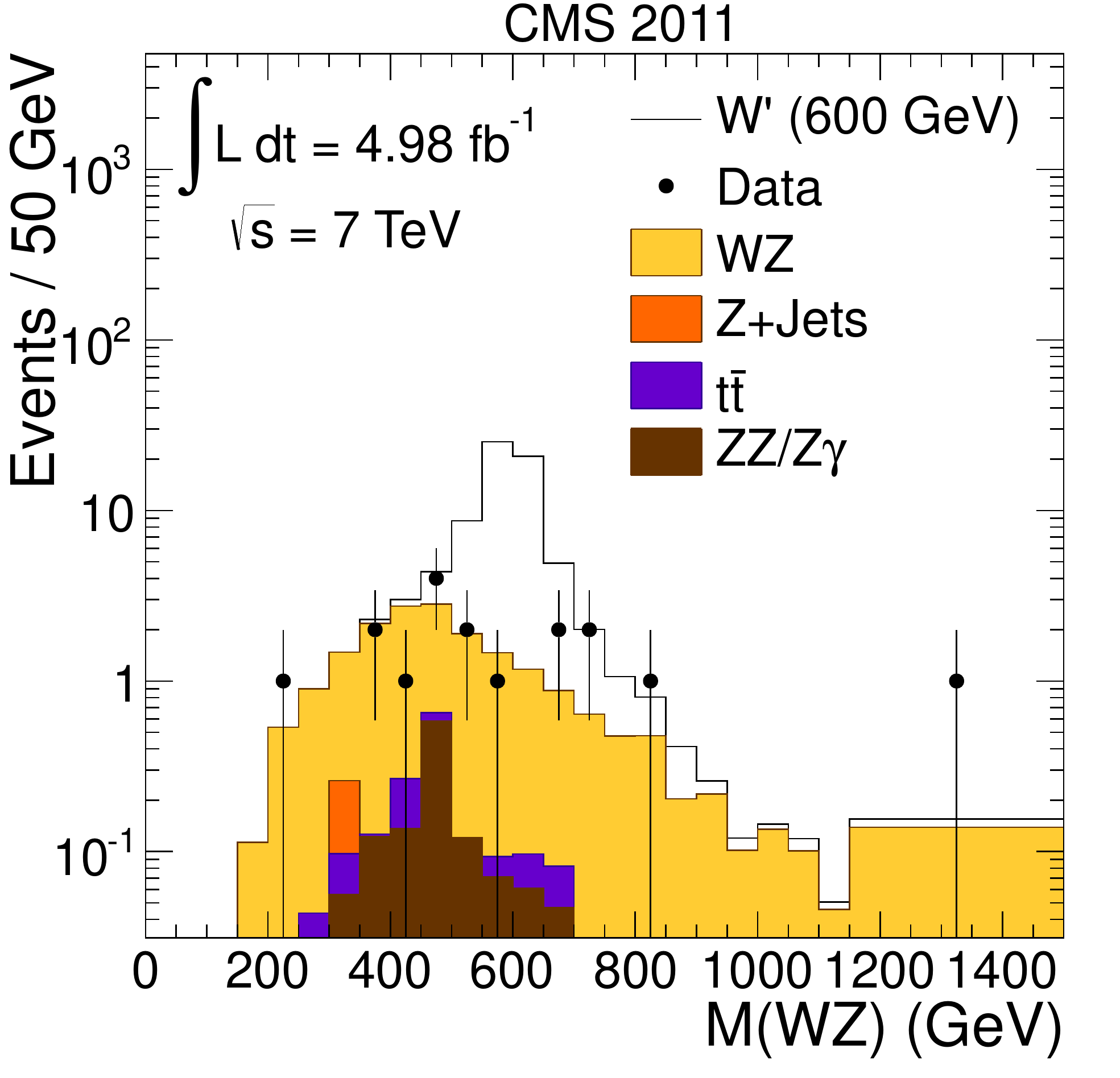}
 \caption{Distribution of the $\PW\cPZ$ invariant mass (\cmsLeft) before application of the $L_\mathrm{T}$ requirement and (\cmsRight) after applying the $L_\mathrm{T}$ requirement ($L_\mathrm{T} > 290$\GeV) for the \PWpr\ mass point at 600\GeV.
}
 \label{WZMass_ValidWZCand}
 \end{center}
 \end{figure}

As the momentum component of the neutrino along the beam direction is unknown, the invariant mass of $\PW\cPZ$ candidates cannot be uniquely determined. However, by assuming the $\PW$ to have its nominal
mass, the value of the neutrino longitudinal momentum is constrained to one of the two solutions of a quadratic equation. Owing to detector resolution effects, the reconstructed
transverse mass was found to lie above the invariant $\PW$ mass, $M(\PW)$, in 20\% of events, leading to complex solutions for the neutrino longitudinal momentum.
In these cases, a real solution was recovered by setting $M(\PW)$ equal to the measured transverse mass. This results in two identical solutions for the neutrino longitudinal momentum. In simulated events with two unique solutions, the smaller-magnitude solution
was found to be correct in approximately 75\% of the cases, and this solution was therefore chosen for all events. The observed invariant mass distribution of $\PW\cPZ$ candidates, $M(\PW\cPZ)$, is shown on the \ifthenelse{\boolean{cms@external}}{upper panel}{left-hand side} of Fig.~\ref{WZMass_ValidWZCand}.

In order to optimise the expected upper limit on the signal cross section, an additional selection requirement was applied on the scalar sum of the transverse momenta of
the charged leptons coming from the W and Z bosons ($L_\mathrm{T}$). For each $\PWpr/\rhoT$ mass hypothesis considered, an optimised $\PW\cPZ$ mass search window and a minimum $L_\mathrm{T}$ requirement were
jointly determined to give the best expected limit. The chosen $L_\mathrm{T}$ and mass-window requirements are listed in Table~\ref{tab:MassWindows}, and the $\PW\cPZ$ invariant
mass after applying the $L_\mathrm{T}$ requirement for the \PWpr\ mass point at 600\GeV is shown on the \ifthenelse{\boolean{cms@external}}{lower panel}{right-hand side} of Fig.~\ref{WZMass_ValidWZCand}. There is no excess observed in the data above the expected standard model background.

As a cross-check of the simulation, the \cPZ+jets and $\ttbar$ backgrounds were estimated from the data by measuring the efficiencies for
genuine and misidentified leptons to pass the isolation criteria, and applying those efficiencies to a sample of events passing all requirements except
for isolation. The total background result agrees with the numbers from simulation, and the uncertainties assigned to the Z+jets and $\ttbar$
contributions when determining limits were based on the uncertainties in the estimates from data.

\label{sec:systematics}

Systematic uncertainties affecting the product of acceptance, reconstruction, and identification efficiencies for the final-state objects were determined from simulation. These include uncertainties stemming from lepton and \MET energy scales and resolutions, NLO effects, and pile-up simulation. Following the recommendations of the PDF4LHC group~\cite{PDF4LHC}, PDF and $\alpha_s$ variations of the MSTW2008~\cite{Martin:2009iq}, CTEQ6.6~\cite{Nadolsky:2008zw} and NNPDF2.0~\cite{Ball:2010de} PDF sets were taken into account and their impact on the signal cross sections estimated.

The uncertainty on the background simulation is dominated by the 10\% uncertainty due to NLO K-factor corrections for the $\PW\cPZ$ component. Cross-section uncertainties of 7.5\% for $\cPZ\cPZ$~\cite{MCFM}, 13\% for $\cPZ\gamma$~\cite{VGammaCMS}, and 17\% for $\PW\cPZ$~\cite{WZAtlas2012} were also taken into account, along with a 2.2\% uncertainty on the integrated luminosity~\cite{LUMIPAS2012}.

\label{sec:Results}

Exclusion limits on the production cross section $\sigma (\Pp\Pp \to \PWpr/\rhoT \rightarrow \PW\cPZ) \times
{\mathcal{B}}(\PW\cPZ \to 3 \ell \Pgn)$ were determined by comparing the number of observed events with the numbers of expected signal and background events in each search window. The calculations were performed using the {\tt RooStats} implementation~\cite{roostats} of the CL$_s$ statistic. The event counts and efficiencies are shown in Table~\ref{tab:MassWindows}. We note that the efficiency drops at high $\PWpr$ mass because of the isolation requirement as the leptons from the boosted $\cPZ$ boson become more collimated.  We interpolate between mass points where we have simulated the signal to establish mass limits for each model.

\begin{figure}[!htb]
 \begin{center}
 \includegraphics[width=0.45\textwidth]{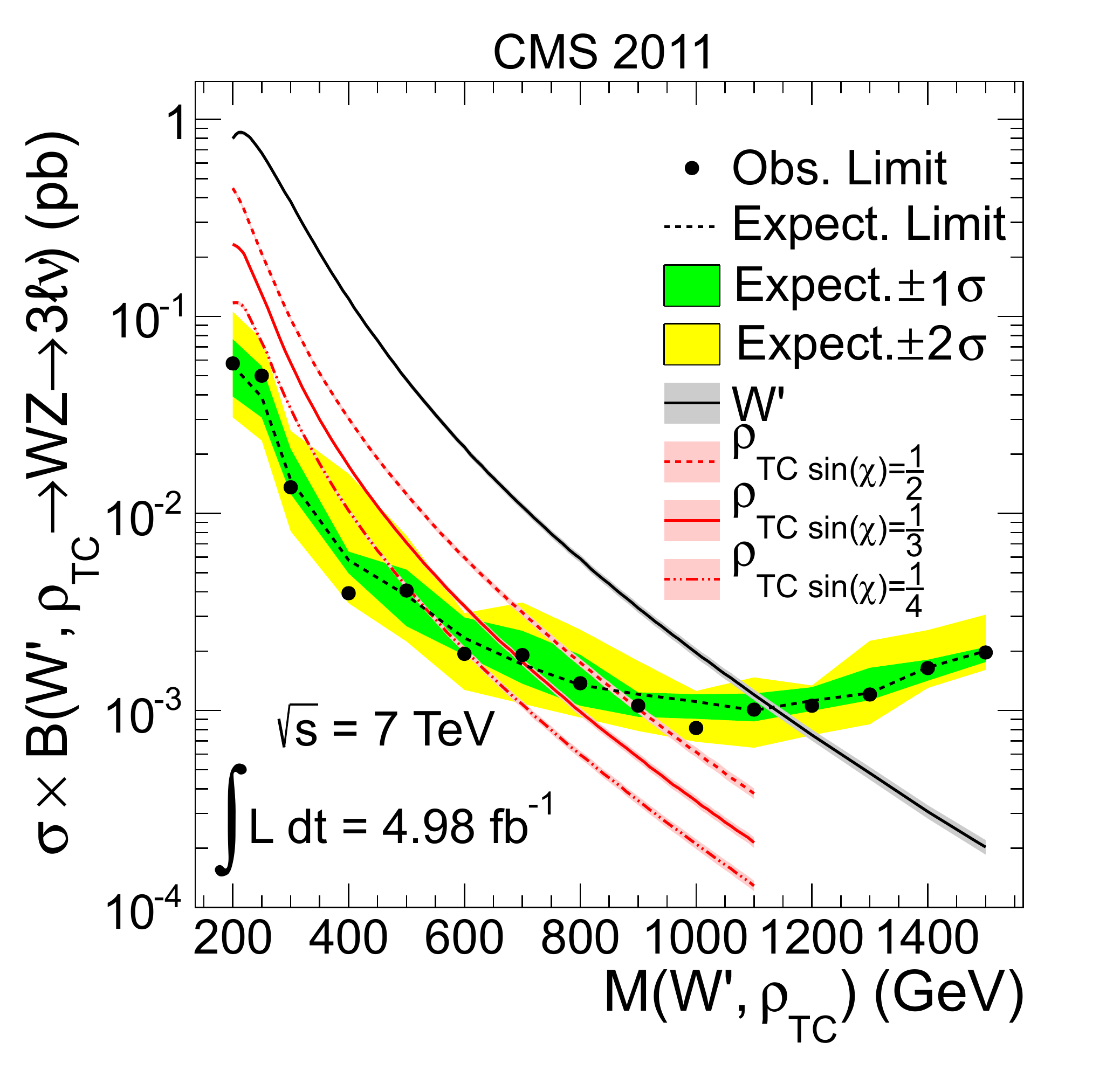}
 \caption{Expected and observed 95\% CL upper limits on $\sigma \times
   {\mathcal{B}}(\PWpr/\rhoT \to 3 \ell \Pgn)$ as a function of the resonance
   mass for $\PWpr$ or $\rhoT$, along with the $1\sigma$ and $2\sigma$
   combined statistical and systematic expected variation shown as a green (dark)
and  yellow (light) band, respectively. The theoretical cross sections, with PDF
   uncertainties, include a mass-dependent NNLO K-factor.
  The LSTC cross sections assume the relationship $M(\piT)=\frac{3}{4}M(\rhoT) - 25 \GeV$.}
 \label{fig:LimitVsMass}
 \end{center}
 \end{figure}

\begin{figure}[!htb]
 \begin{center}
 \includegraphics[width=0.45\textwidth]{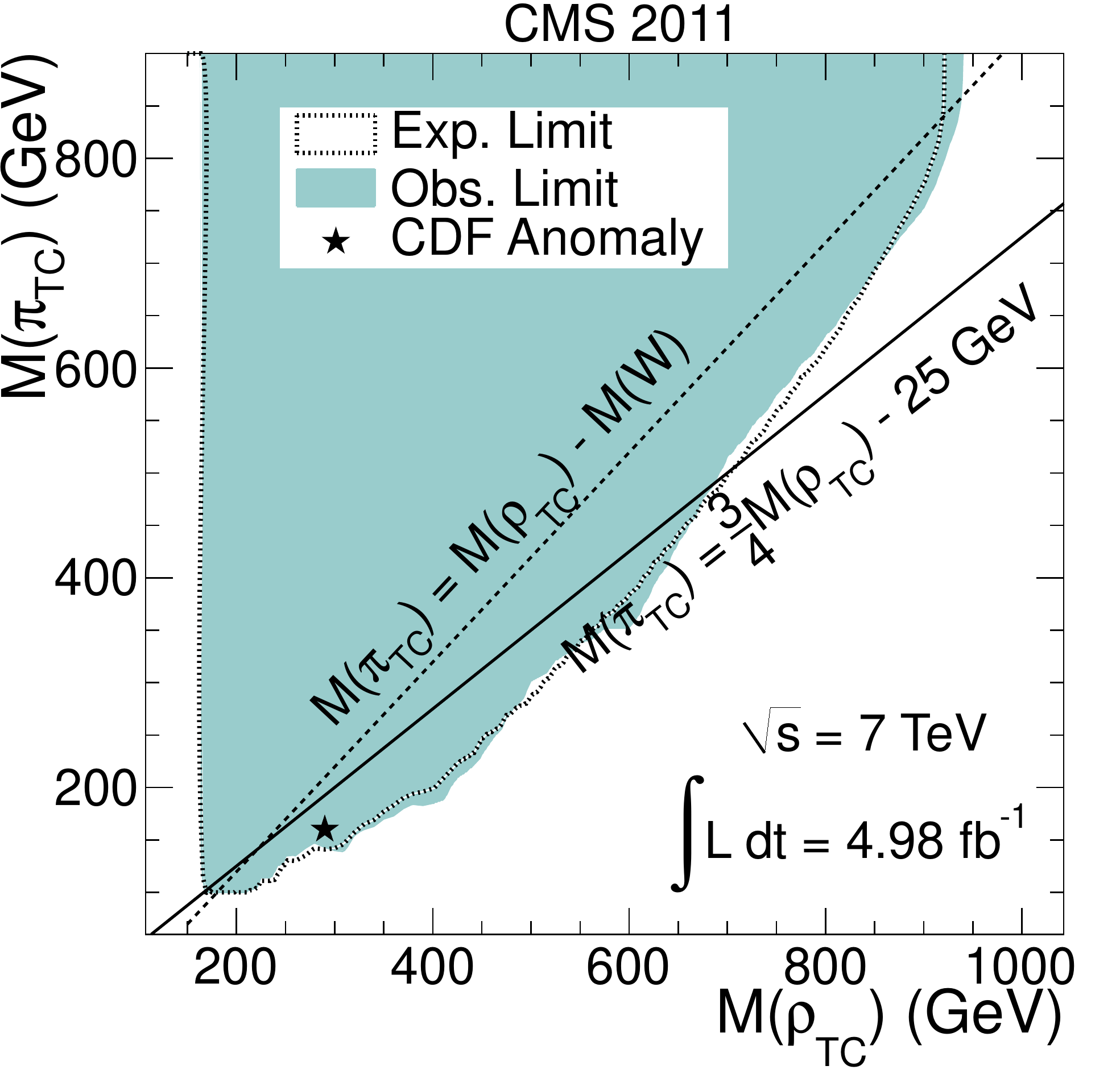}
 \caption{Exclusion region at 95\% CL for LSTC as a function of the $\rhoT$ and $\piT$ masses.}
 \label{fig:2Dlimit}
 \end{center}
 \end{figure}

In the SSM, these limits allow the exclusion of $\PWpr$ bosons with masses below \wprimeLimit\GeV (Fig.~\ref{fig:LimitVsMass}).
For LSTC, with the chosen parameters $M(\piT)=\frac{3}{4}M(\rhoT) - 25\GeV$, $\rhoT$ hadrons with masses between \tcLesHouchesLimitLower\space and \tcLesHouchesLimit\GeV are excluded (see Fig.~\ref{fig:2Dlimit}). Figure~\ref{fig:2Dlimit} also shows LSTC limits determined as a function of the $\rhoT$ and $\piT$ masses. The lower mass limits are obtained by extrapolating below 200\GeV. For the parameters  chosen by the \DZERO experiment, $M(\rhoT) < M(\piT) + M(\PW)$, more stringent limits are obtained, excluding the range \tcOptimisticLimitObsLower\space to \tcOptimisticLimitObs\GeV for $M(\rhoT)$. It can be seen that the LSTC interpretation of a deviation from the SM observed by the CDF experiment in the \PW+jets channel~\cite{Aaltonen:2011mk}, with proposed parameters $M(\rhoT) = 290\GeV$, $M(\piT) = 160\GeV$~\cite{PhysRevLett.106.251803}, is excluded by the 95\% CL upper bound of 150\GeV on $M(\piT)$ for the required \rhoT mass. A more recent publication~\cite{Eichten:2012br} proposes the evaluation of the cross section for
$\rhoT \to \PW\cPZ$ as a function of the LSTC parameter $\sin\chi$. Changes in this parameter impact the decay branching fractions for $\PW\cPZ$ and $\PW\piT$ (the channel studied in Ref.~\cite{PhysRevLett.106.251803}), among others. Figure~\ref{fig:LimitVsMass} shows the predicted cross sections corresponding to $\sin{\chi}=\frac{1}{2}$ and $\sin{\chi}=\frac{1}{4}$, as well as the value $\sin{\chi} = \frac{1}{3}$ used to establish the reported limit.

\begin{figure}[!htb]
 \begin{center}
 \includegraphics[width=0.45\textwidth]{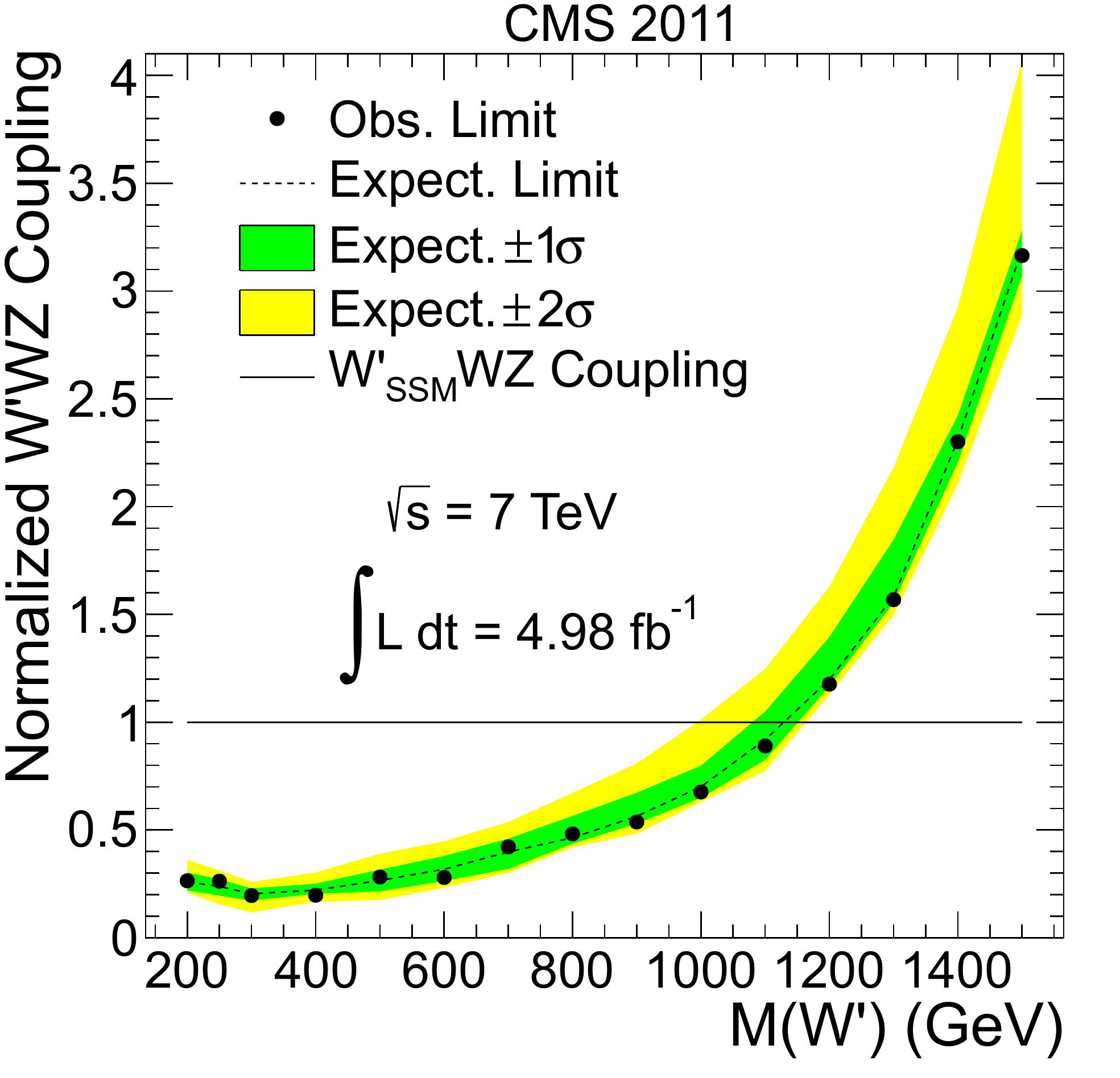}
 \caption{Upper limit at 95\% CL on the strength of the $\PWpr$ $\PW\cPZ$ coupling
normalized to the SSM prediction, as a function of the $\PWpr$ mass. The $1\sigma$ and $2\sigma$ combined statistical and systematic expected variation is shown as a green (dark) and yellow (light) band, respectively. PDF uncertainties on the theoretical cross section are not included.}
 \label{fig:wprimewz_limit}
 \end{center}
 \end{figure}

The $\PWpr$ production cross section and the $\PWpr \to \PW\cPZ$ branching fraction are both affected by the strength of the coupling between the $\PWpr$, $\PW$ and $\cPZ$ bosons. Figure~\ref{fig:wprimewz_limit} shows the 95\% CL upper limit on the coupling as a function of the mass of the $\PWpr$ resonance.

\begin{table}[!h]
\topcaption{Minimum $L_\mathrm{T}$ requirements and search windows for each $\PWpr$/(\rhoT) mass point along with the number of expected background events, observed events, and signal efficiency. Indicated uncertainties are statistical and systematic combined in quadrature.}
\begin{center}
\begin{tabular}{ccr@{--}lr@{$\pm$}lcr@{$\pm$}l }
  \hline
  $M(\PWpr/\rhoT)$ & $L_\mathrm{T}$ & \multicolumn{2}{c}{$\PW\cPZ$}  & \multicolumn{2}{c}{$N_\text{Bkg}^\mathrm{MC}$} & Data & \multicolumn{2}{c}{$\epsilon_\text{Sig}$} \\
            &  & \multicolumn{2}{c}{Mass Window}   &   \multicolumn{2}{c}{}     &                & \multicolumn{2}{c}{}                  \\ (\GeVns) & (\GeVns) & \multicolumn{2}{c}{(\GeVns)}   &   \multicolumn{2}{c}{}     &                & \multicolumn{2}{c}{(\%)}                  \\ \hline       \hline
 200 &   0 & 190&210 & 50 & 9 & 52 & 8.0 & 0.4 \\
 250 & 150 & 230&270 & 34 & 6 & 40 & 8.8 & 0.4 \\
 300 & 160 & 280&320 & 24 & 5 & 23 & 18 & 1 \\
 400 & 220 & 360&440 & 13 & 2 & 7 & 29 & 1 \\
 500 & 230 & 450&550 & 8 & 2 & 9 & 41 & 1 \\
 600 & 290 & 540&660 & 3.4 & 0.7 & 2 & 45 & 1 \\
 700 & 360 & 620&780 & 1.8 & 0.4 & 2 & 48 & 1 \\
 800 & 400 & 710&890 & 1.0 & 0.2 & 1 & 52 & 2 \\
 900 & 400 & 760&1040 & 1.0 & 0.2 & 0 & 61 & 2 \\
 1000 & 400 & 820&1180 & 0.8 & 0.2 & 0 & 65 & 2 \\
 1100 & 400 & 890&1310 & 0.6 & 0.1 & 0 & 63 & 1 \\
 1200 & 400 & 940&1460 & 0.4 & 0.1 & 0 & 58 & 1 \\
 1300 & 400 & 1020&1580 & 0.3 & 0.1 & 0 & 50 & 1 \\
 1400 & 400 & 1110&1690 & 0.18 & 0.05 & 0 & 36 & 1 \\
 1500 & 400 & 1200&1800 & 0.13 & 0.04 & 0 & 30 & 1 \\ \hline
\end{tabular}
\label{tab:MassWindows}
\end{center}
\end{table}

In summary, a search for new exotic particles decaying via $\PW\cPZ$ to final
states with electrons and muons has been performed using pp collisions at $\sqrt{s}=7$\TeV corresponding to an integrated luminosity of \usedLumiWZ, collected by the CMS experiment. No significant excess was observed in the
invariant mass distribution of $\PW\cPZ$ candidates, compared to the expectation
from standard model processes. The results have been interpreted in the
context of several theoretical models, with the data used to establish bounds at 95\% CL on the masses of hypothetical particles decaying
via $\PW\cPZ$. In the framework of the sequential standard model, $\PWpr$ bosons
with masses below \wprimeLimit\GeV have been excluded. Technicolor $\rhoT$ hadrons
with masses between \tcLesHouchesLimitLower\space and \tcLesHouchesLimit\GeV have been
excluded, assuming $M(\piT)=\frac{3}{4}M(\rhoT) - 25\GeV$. Under the
alternative assumption $M(\rhoT) < M(\piT) + M(\text{W})$, $\rhoT$ hadrons with
masses between \tcOptimisticLimitObsLower\space and \tcOptimisticLimitObs\GeV have been
excluded. These are the most stringent limits to date in the $\PW\cPZ$ channel.

We thank Kenneth Lane for his help with the interpretation of the results within the context of low-scale technicolor models. We wish to congratulate our colleagues in the CERN accelerator departments for the excellent performance of the LHC machine. We thank the technical and administrative staff at CERN and other CMS institutes, and acknowledge support from: FMSR (Austria); FNRS and FWO (Belgium); CNPq, CAPES, FAPERJ, and FAPESP (Brazil); MES (Bulgaria); CERN; CAS, MoST, and NSFC (China); COLCIENCIAS (Colombia); MSES (Croatia); RPF (Cyprus); Academy of Sciences and NICPB (Estonia); Academy of Finland, MEC, and HIP (Finland); CEA and CNRS/IN2P3 (France); BMBF, DFG, and HGF (Germany); GSRT (Greece); OTKA and NKTH (Hungary); DAE and DST (India); IPM (Iran); SFI (Ireland); INFN (Italy); NRF and WCU (Korea); LAS (Lithuania); CINVESTAV, CONACYT, SEP, and UASLP-FAI (Mexico); MSI (New Zealand); PAEC (Pakistan); SCSR (Poland); FCT (Portugal); JINR (Armenia, Belarus, Georgia, Ukraine, Uzbekistan); MON, RosAtom, RAS and RFBR (Russia); MSTD (Serbia); MICINN and CPAN (Spain); Swiss Funding Agencies (Switzerland); NSC (Taipei); TUBITAK and TAEK (Turkey); STFC (United Kingdom); DOE and NSF (USA).

\bibliography{auto_generated}   

\cleardoublepage \appendix\section{The CMS Collaboration \label{app:collab}}\begin{sloppypar}\hyphenpenalty=5000\widowpenalty=500\clubpenalty=5000\textbf{Yerevan Physics Institute,  Yerevan,  Armenia}\\*[0pt]
S.~Chatrchyan, V.~Khachatryan, A.M.~Sirunyan, A.~Tumasyan
\vskip\cmsinstskip
\textbf{Institut f\"{u}r Hochenergiephysik der OeAW,  Wien,  Austria}\\*[0pt]
W.~Adam, E.~Aguilo, T.~Bergauer, M.~Dragicevic, J.~Er\"{o}, C.~Fabjan\cmsAuthorMark{1}, M.~Friedl, R.~Fr\"{u}hwirth\cmsAuthorMark{1}, V.M.~Ghete, J.~Hammer, N.~H\"{o}rmann, J.~Hrubec, M.~Jeitler\cmsAuthorMark{1}, W.~Kiesenhofer, V.~Kn\"{u}nz, M.~Krammer\cmsAuthorMark{1}, D.~Liko, I.~Mikulec, M.~Pernicka$^{\textrm{\dag}}$, B.~Rahbaran, C.~Rohringer, H.~Rohringer, R.~Sch\"{o}fbeck, J.~Strauss, A.~Taurok, W.~Waltenberger, G.~Walzel, E.~Widl, C.-E.~Wulz\cmsAuthorMark{1}
% ...
\vskip\cmsinstskip
\dag:~Deceased\\
1:~~Also at Vienna University of Technology, Vienna, Austria\\
% ...

\end{sloppypar}
\end{document}